\documentstyle[prl,preprint,epsf,aps]{revtex}
\begin{document}
\draft
\title{A New Type of Intensity Correlation in Random Media}
\author{B. Shapiro}
\address{Department of Physics, Technion-Israel Institute
of Technology, Haifa 32000, Israel.}
\maketitle
\begin{abstract}
A monochromatic point source, embedded in a three-dimensional
disordered medium, is considered.  The resulting intensity pattern
exhibits a new type of long-range correlations. The range of these
correlations is infinite and their magnitude, normalized to the
average intensity, is of order $1/k_0 \ell$, where $k_0$ and $\ell$
are the wave number and the mean free path respectively.
\end{abstract}

A wave propagating in a random medium undergoes multiple scattering and
produces a complicated, irregular intensity pattern. Such a pattern is
described in statistical terms and one of its important characteristics
is the correlation function
$\langle I(\vec{r}_1) I(\vec{r}_2)\rangle  $,
where
$I(\vec{r})$
is the wave intensity at point
$\vec{r}$
and the angular brackets designate the ensemble average. This correlation
function has been studied in recent years, both theoretically and
experimentally, and various types of correlations has been
identified$^1$.  There are large short-range correlations, due to
interference among the waves arriving, via all possible scattering
sequences, to a neighborhood of a given point; this interference is
responsible for the large spatial fluctuations in intensity
(speckles). There exist also weak long-range correlations, due to
diffusion which propagates the locally large fluctuations to distant
regions in space. The purpose of this note is to identify and study
a new type of long-range correlation which, under appropriate
circumstances, can dominate  the ``standard''
long-range correlations$^1$.

I consider a monochromatic point source embedded in an infinite
three-dimensional random medium. The position of the source is
$\vec{r}_0$ and its frequency is $\omega$.  Assuming a scalar wave,
one has the following equation for the field (the Green's function)
at point $\vec{r}$:
\begin{eqnarray}
\left\{ \nabla^2 +k_0^2 [1 + \mu (\vec{r})] + i \eta \right\}
G_{\omega} (\vec{r}, \vec{r}_0) =
\delta (\vec{r} -  \vec{r}_0) \ .    
\end{eqnarray}
Here $\mu (\vec{r})$
is the fractional fluctuation of the dielectric constant,
$\eta$ is a positive
infinitesimal and
$k_0 = \omega/c$,
where $c$ is the speed of propagation in the average medium.  I assume
that $\mu(\vec{r})$ obeys white-noise Gaussian statistics, i.e.
\begin{eqnarray}
\langle \mu (\vec{r}) \mu(\vec{r}^{\prime})\rangle   =
u \delta (\vec{r} - \vec{r}^{\prime}) \ ,    
\end{eqnarray}
where the constant $u$ describes the strength of the disorder.

For weak disorder, the average field, the average intensity and various
correlation functions can be computed by the diagram technique$^1$.
The average field,
$\langle G_{\omega} (\vec{r}, \vec{r}_0)\rangle  $,
decays exponentially away from the source:
\begin{eqnarray}
\langle G_{\omega} (\vec{r}, \vec{r}_0)\rangle   = -
\frac{1}{4\pi \mid \vec{r} - \vec{r}_0 \mid}
\exp \left[ (ik_0 - \frac{1}{2\ell})
\mid \vec{r} - \vec{r}_0 \mid \right] \ ,   
\end{eqnarray}
where
$\ell = 4\pi/uk^4_0$
is the mean free path and
$k_0 \ell \gg 1$.

The intensity at point $\vec{r}$ is defined as
$I_{\omega}(\vec{r}) = \mid G_{\omega} (\vec{r}, \vec{r}_0) \mid^2$
and its average value is given by the diagram in Fig.~1. Intensity
propagates from the source to a distant observation point, such that
$\mid \vec{r} - \vec{r}_0 \mid \gg \ell$, by a diffusion process. This
is represented by a diffusion ladder (the shaded box),
$T (\vec{r}_1, \vec{r}_2) = 3/\ell^3
\mid \vec{r}_1 - \vec{r}_2 \mid$.
The vertices, connecting the external points to the ladder, are
short-range objects which can be replaced by the number
\begin{eqnarray}
\int d^3 r_1 \langle G_{\omega} (\vec{r}_0, \vec{r}_1) \rangle
\langle G^{\ast}_{\omega} (\vec{r}_0, \vec{r}_1) \rangle
= \frac{\ell}{4\pi} \ ,     
\end{eqnarray}
so that the average intensity at point $\vec{r}$ is
\begin{eqnarray}
\langle I_{\omega}(\vec{r}) \rangle   =
\left( \frac{\ell}{4\pi}\right)^2
T(\vec{r}, \vec{r}_0) =
\frac{3}{16\pi^2 \ell \mid \vec{r} - \vec{r}_0 \mid} \ .  
\end{eqnarray}

Let us now consider the correlation function
\begin{eqnarray}
\frac{\langle \Delta I_{\omega}(\vec{r}) \Delta I_{\omega} (\vec{r} +
\Delta \vec{r}) \rangle  }{\langle I_{\omega}(\vec{r})\rangle
\langle I_{\omega} (\vec{r} +
\Delta \vec{r}) \rangle  } \equiv C(\Delta r) \ .    
\end{eqnarray}
where
$\Delta I_{\omega}(\vec{r}) = I_{\omega}(\vec{r}) -
\langle I_{\omega} (\vec{r}) \rangle  $
is the deviation from the average value. The points $\vec{r}$ and
$\vec{r} + \Delta \vec{r}$ are assumed to be far from the source, i.e.,
many mean free paths away. Since only long-range correlations are
considered here, $\Delta r \gg \ell$ is assumed.

The leading contribution to the long-range correlations is represented
by the diagram in Fig.~2. The diagram contains two pairs of Green's
functions. One pair propagates, via a diffusion ladder, to point
$\vec{r}$; \ the other - to point
$\vec{r} + \Delta \vec{r}$.
The pairs are connected by a dashed line, which makes this diagram
belong to the set of diagrams contributing to $C(\Delta r)$. The
physical interpretation of this diagram is straightforward: \ the wave,
emanating from the source, gets scattered at some point $\vec{r}_1$,
close to the source. The secondary wave, emerging from $\vec{r}_1$,
propagates by diffusion to two distant points, $\vec{r}$ and
$\vec{r} + \Delta \vec{r}$.
In this way intensity correlation at the two points is established.

In order to evaluate the diagram in Fig.~2 one has to compute the
vertex, depicted separately in Fig.~3. The vertex is a short-range
object and, thus, can be replaced by a point, located at $\vec{r}_0$,
from which the two diffusion ladders emerge. The number $V$, assigned
to the point is:
\begin{eqnarray}
V &=& \frac{4\pi}{\ell} \int
d^3r_1 d^3r_2 d^3r_3 \langle G_{\omega}(\vec{r}_2 - \vec{r}_0)\rangle 
\langle G^{\ast}_{\omega}(\vec{r}_2 - \vec{r}_1)\rangle
\langle G^{\ast}_{\omega}(\vec{r}_1 - \vec{r}_0)\rangle\times \nonumber
\\
&\ \ \ & \times \langle G_{\omega}(\vec{r}_3 - \vec{r}_1)\rangle
\langle G_{\omega}(\vec{r}_1 - \vec{r}_0)\rangle
\langle G^{\ast}_{\omega}(\vec{r}_3 - \vec{r}_0)\rangle   \ .   
\end{eqnarray}
Integration over $\vec{r}_2$ gives
\begin{eqnarray}
\int d^3r_2 \langle G_{\omega}(\vec{r}_2 - \vec{r}_0)\rangle
\langle G^{\ast}_{\omega}(\vec{r}_2 - \vec{r}_1)\rangle
= \frac{\ell}{4\pi}
f_{\omega} ( \mid \vec{r}_0 - \vec{r}_1 \mid ) \ .  
\end{eqnarray}
where
\begin{eqnarray}
f_{\omega}(x) = \frac{\sin k_0 x}{k_0 x}
\exp \left(- \frac{x}{2 \ell}\right)  \ .        
\end{eqnarray}
Integration over $\vec{r}_3$ gives an identical contribution, so that
\begin{eqnarray}
V &=& \frac{4 \pi}{\ell}
\left( \frac{\ell}{4\pi} \right)^2
\int d^3r_1
\langle G_{\omega}(\vec{r}_1 - \vec{r}_0)\rangle
\langle G^{\ast}_{\omega}(\vec{r}_1 - \vec{r}_0)\rangle
f^2_{\omega} (\mid \vec{r}_0 - \vec{r}_1 \mid) = \nonumber \\
&=& \frac{\ell}{4\pi}
\int d^3 \rho \frac{1}{(4\pi \rho)^2}
e^{-2\rho /\ell}
\left( \frac{\sin k_0 \rho}{k_0 \rho} \right)^2  \ .  
\end{eqnarray}
The integral is dominated by the region of small $\rho$, i.e.
$\rho \simeq k^{-1}_0 \ll \ell$,
so that the exponential can be replaced by 1, which leads to
$V = \ell /32 \pi k_0$.
The diagram in Fig.~2 is, thus, equal to
\begin{eqnarray}
\frac{\ell}{32\pi k_0} T (\vec{r}_0, \vec{r})
T (\vec{r}_0, \vec{r} + \Delta \vec{r})
\left( \frac{\ell}{4\pi} \right)^2           
\end{eqnarray}
where the factor
$\left( \frac{\ell}{4\pi} \right)^2$
accounts for the two vertices connecting the diffusion ladders to the
observation points $\vec{r}$ and
$\vec{r} + \Delta \vec{r}$.
Recalling that
$\langle I_{\omega}(\vec{r})\rangle   =
(\ell /4\pi)^2
T(\vec{r}, \vec{r}_0)$,
and similarly for
$\langle I_{\omega}(\vec{r} + \Delta \vec{r})\rangle$,
one can rewrite Eq.~(11) as
$(\pi /2k_0 \ell) \langle I_{\omega}(\vec{r})\rangle
\langle I_{\omega}(\vec{r} + \Delta \vec{r})\rangle$.
Finally, the diagram in Fig.~2 should be assigned a combinatorial
factor 2, since one could connect the two external Green's function
by a dashed line, instead of connecting the two internal ones as in
Fig.~2. (Only lines with arrows in opposite directions, corresponding
to a pair of $G,G^{\ast}$ need to be connected.) \ Thus, the contribution
of the diagram in Fig.~2 to the normalized correlation function
[Eq.~(6)] is:
\begin{eqnarray}
C_0 (\Delta r) = \frac{\pi}{k_0 \ell}  
\end{eqnarray}
where the notation $C_0$ is used to distinguish this term from the other
types of correlations, usually designated$^1$ as $C_1$, $C_2$ and
$C_3$. For the  geometry considered, i.e. point source in an infinite
three-dimensional medium, the $C_0$-term is much larger than the other
types of long-range correlations.  Moreover, it does not decay in space
(infinite-range correlation).

It is worth to note that the $C_0$-term resembles the $C_2$-term$^{1,2}$,
in the sense that both terms descibe correlation between two distant
points, for intensity propagating from a single source. However, $C_0$ 
is not contained in $C_2$. Indeed, $C_2$ is desribed by a 4-ladder
diagram and, thus, cannot contain the diagram in Fig.~2. Moreover, in
computing and interpreting the $C_2$-correlation it is essential that
all 4 ladders are "at work" and can be treated in the diffusion
 approximation$^{1,2}$. For a point source in three dimensions
the $C_2$-term is of order $(k_0\ell)^{-2}$, which is much smaller
than the $C_0$-term in Eq.~(12).

So far $\Delta r \gg\ell$ was assumed. The case $\Delta r = 0$, however,
is also of interest. For this case the diagram in Fig. 2 gives a
contribution to the second moment of intensity. For the normalized
intensity,
$\tilde{I}(\vec{r})\equiv I(\vec{r})/\langle I(\vec{r})\rangle $,
this contribution is $2C_0(0) = 2\pi/k_0 \ell$
(the extra factor 2, as compared to Eq. (12), appears because for
$\Delta r = 0$ one can pair also Green's functions emerging from point
$\vec{r}$). This contribution represents a small correction to the
Rayleigh value $\langle \tilde{I}^2\rangle   = 2$.  The knowledge of the correction
enables one to compute deviations of the intensity distribution
$P(\tilde{I})$ from the Rayleigh function
$P_0 (\tilde{I}) = \exp (-\tilde{I})$.
The calculation is practically identical to the one performed by
Mirlin et al.$^3$, with one important difference: Ref.~3 considered
quasi-one dimensional geometry, where the correction was quite different
from the present value $2\pi /k_0 \ell$.  For the present,
three-dimensional case the result is:
\begin{eqnarray}
P (\tilde{I}) \simeq \exp \left(- \tilde{I} +
\frac{\pi}{2k_0 \ell} \tilde{I}^2\right) \ , \ \ \ \ \
\tilde{I} \ll k_0 \ell  \ .      
\end{eqnarray}
The asymptotic tail of the distribution, for
$\tilde{I} \gg k_0 \ell$,
can be obtained by the method of optimal fluctuation$^4$. For
$\tilde{I} \simeq k_0 \ell$
the tail should smoothly match expression~(13). A detailed
calculation of $P(\tilde{I})$ will be presented in a separate
publication.

One can easily extend the calculation to include correlations
at different frequencies. The object to consider is the same as in
Eq.~(6), but with
$I_{\omega + \Delta \omega} (\vec{r} + \Delta \vec{r})$
instead of
$I_{\omega} (\vec{r} + \Delta \vec{r})$.
The only difference in the calculation is that, in Eq.~(7) for $V$,
one has to substitute $\omega + \Delta \omega$ for $\omega$ in the last
three Green's
functions. This leads to a replacement $\omega \rightarrow \omega
+ \Delta \omega$ in one of the Green's function and in one of the
f-factors in Eq.~(10). For $\Delta \omega \ll\omega$ this replacement
leads only to inessential correction to the constant value in Eq.~(12).

In conclusion:

\noindent (i) It has been shown that the intensity pattern,
produced by a wave propagating in a random medium, can exhibit a new
type of correlation, designated as $C_0$. These correlation is  of
order $1/k_0 \ell$ and has  infinite range in space and frequency. 
Unlike the other types of
long-range correlations, $C_2$ and $C_3$, the $C_0$-term is dominated
by the configuration of the disorder near the source.

\noindent (ii) Being sensitive to the short-distance properties of the
disorder, the $C_0$-term cannot be universal, i.e. its magnitude should
depend
on the specific type of the disorder. For the white-noise Gaussian
disorder, considered in this note, $C_0$ is determined by the parameter
$k_0 \ell$. However, for Gaussian disorder with a finite correlation
length $d$, or for discrete impurities of size $d$, the $C_0$-term should
depend on $d$.

\noindent (iii) Let us emphasize that this paper considers a propagating
wave, in an open system. A somewhat different problem pertains to
the statistics of eigenmodes, and eigenfrequencies, in a closed disordered
system. Non-universal terms, of the same origin as these considered here, exist
also in that case, as has been recently discussed by Mirlin$^5$.

\noindent (iv) The $C_0$ term is also present in lower dimensionalities.
The quasi-one dimensional geometry (a tube of length $L$ and cross
section $A$) is of a particular interest, since it is often used in
experiments$^6$. The $C_2$ and $C_3$-terms in this geometry are of
order $g^{-1}$ and $g^{-2}$, respectively, where
$g \simeq k^2_0 A \ell/L$. Thus, for a point source inside the tube,
the $C_0$ term will dominate as long as $k_0 \ell \leq g$, i.e.
$L < A k_0$. This implies that our earlier  calculation$^3$,
which left out the $C_0$-term, is valid only for $L>Ak_0$.

I acknowledge previous collaboration with A. Mirlin and R. Pnini.
It paved the way for the present work.  Useful conversations with
A. Genack, D. Khmelnitskii and A. Lagendijk are gratefully acknowledged. The work
was supported by the US-Israel Binational Science Foundation and by the
Fund for promotion of research at the Technion.

\begin{figure}
\epsfysize=2.5in
\centerline{\epsffile{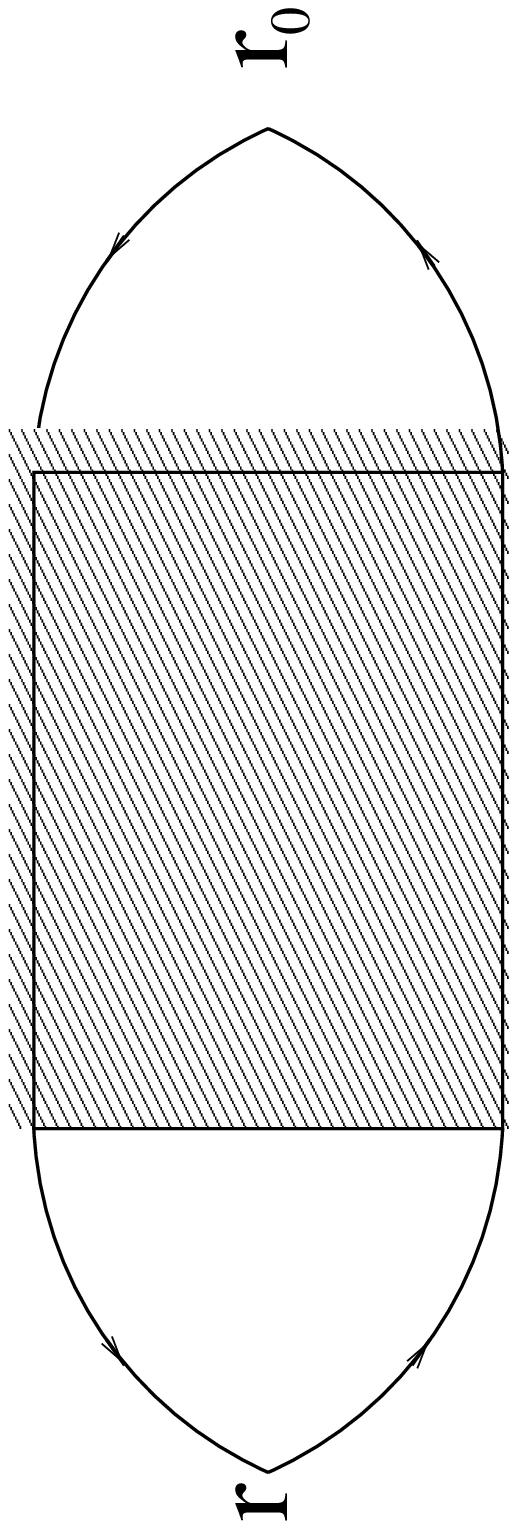}}
\caption{The average intensity.}
\end{figure}

\begin{figure}
\epsfysize=2.5in
\centerline{\epsffile{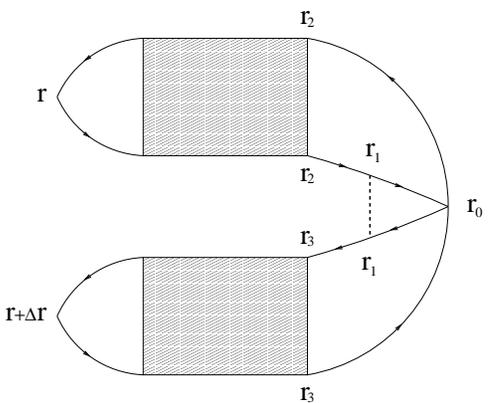}}
\caption{The correlation
function ($C_0$-term).}
\end{figure}

\begin{figure}
\epsfysize=2.0in
\centerline{\epsffile{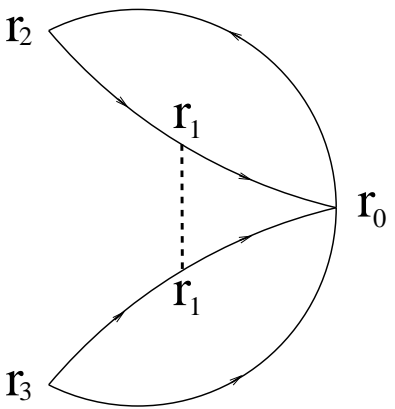}}
\caption{The vertex.}
\end{figure}

\end{document}